\documentclass[journal=jacsat,manuscript=article]{achemso}
\usepackage[version=3]{mhchem} 
\usepackage{graphicx}
\usepackage{xcolor}

\author{Marcel Rudolf}
\author{Andreas Zumbusch}
\email{andreas.zumbusch@uni-konstanz.de}
\phone{+49 (0)7531 882027}
\affiliation[Universität Konstanz]
{Department of Chemistry, Universität Konstanz, Konstanz, Germany}

\title[]{Temporal evolution of interparticle potentials of PMMA colloids in CHB/decaline}

\abbreviations{}

\begin{document}








\begin{abstract}
Colloidal dispersions composed of polymethylmetacrylate particles dispersed in a mixture of cyclohexylbromide and decalin find widespread use as model systems in optical microscopy experiments. While the system allows simultaneous density and refractive index matching, preparing particles with hard potentials remains challenging and strong variations in the physical parameters of samples prepared in the same manner are common. Here, we present data on the measurement of forces between individual pairs of particles over the course of tens of days using a blinking optical tweezer method. Our results show that the variations in the particle properties are indeed caused by a temporal evolution of the particles' charging. Additional measurements of the influence of tetrabutylammonium bromide (TBAB) addition to the dispersions show that already small concentrations of added TBAB drastically decrease the electrostatic forces between colloidal particles. However, small, non-negligible contact potentials remain even at the highest TBAB concentrations added. 
\end{abstract}



\maketitle

\section{Introduction}\label{sec1}

Apart from their importance in nature and technology, colloidal particles are attractive models for atomic and molecular systems \cite{Manoharan2015}. With typical diameters in the micrometer range, they are easily visualized e.g. using confocal fluorescence microscopy \cite{vanBlaaderen1995,Crocker1996}. In such optical experiments, colloidal particles can be considered as 'big atoms' that are detectable on an individual particle basis \cite{Poon2004}. The tracking of thousands of individual particles gives information that is not accessible in atomic or molecular samples \cite{Weeks2000}. A prototypical example for a phenomenon that can be studied in this manner is the glass transition \cite{Weeks2017}. 

To use colloids as models in optical microscopy experiments, a number of prerequisites has to be fulfilled. In most cases one needs to simultaneously match the density and the refractive index of the colloidal particles to the dispersion medium used: Density matching has the advantage of reducing the movement of particles due to sedimentation during measurements. Refractive index matching of particles and fluid, by contrast, reduces scattering, thus allowing imaging of the particles deep in the sample, and reduces van~der~Waals forces between particles. To date, a few combinations of particles and dispersion media are known for which density and refractive index can be matched at the same time \cite{Wiederseiner2011,Kodger2015,Park2018,Liu2019}. A widely used combination consists of sterically stabilized polymethylmetacrylate (PMMA) colloids in a solution of cyclohexylbromide (CHB) and \textit{cis}-decalin \cite{Yethiraj2003,Royall2003,Kaufman2006,Edmond2012,Besseling2012,Wood2018,Roller2021}. Apart from the two physical properties just discussed, a lot of effort is often taken to prepare particles with hard sphere potentials, i.e. an infinitely high potential for distances smaller than the particle diameter and zero potential everywhere else. Particles of this type are attractive due to their physical simplicity, since for suspensions of particles with perfectly hard potentials, volume fraction is the only parameter controlling structure and dynamics in dense suspensions and no interparticle forces have to be considered in their theoretical description. However, due to charging, colloids don't possess perfectly hard potentials \cite{Royall2013}. In addition, strong variations of parameters like Debye lengths, particle $\zeta$ potentials, and particle charges are commonly observed even for samples prepared in the same manner \cite{vanderLinden2015}. Since the electrostatic properties of dispersed colloids are known to be very susceptible to impurities such as ions or glue residues introduced by the sample chamber preparation \cite{Choi2019}, one usually assumes imperfections in the preparation as causes for the observed variations. 

Particles with potentials closely resembling hard spheres can be prepared, when the two main contributions to the potential are screened. These contributions are  electrostatic and van~der~Waals forces. To overcome the attractive forces arising from van der Waals forces, frequently a steric stabilizer consisting of poly(12-hydroxystearic acid) (PHSA) is covalently attached to the particles. While the chains of the stabilizer create a certain softness in the potential, surface force measurements proved that the influence of the stabilizer on the potential is short range and not measurable for distances greater 20~nm away from the particles' surfaces \cite{Bryant2002}. Thus, especially for spheres with diameters in the $\mu m$ range, the stabilizer has only little effect on the hardness of the potential. The main deviation from hard potentials therefore arises from electrostatic forces. Several approaches have been used to gain information on electrostatic particle potentials. Indirect information about the potential between pairs of particles has been obtained in different manners: by comparing the radial distribution function of a particle ensemble with results of its theoretical description or simulation results \cite{Hsu2005,Royall2006,Espinosa2010}, by measuring the conductivity and mobility with electrophoresis \cite{Yethiraj2003,Hsu2005,Espinosa2010,Kim2014,vanderLinden2015,Farrokhbin2019}, and by examination of the crystallisation behavior of dispersions as a function of volume fraction \cite{Pusey1986,Dinsmore2001,Yethiraj2003,Royall2006}. Direct measurements of interparticle forces are possible using optical tweezers to trap particles while monitoring their interaction via distance measurements. Variations of this approach have been used to study interparticle potentials in a number of different systems \cite{Sainis2007,Sainis2008,Sainis2008a,Gutsche2011,ElMasri2011,Evans2016,Choi2019,Liu2019}.

Here, we report quantitative measurements of interparticle forces based on a method known as blinking optical tweezers, first established by Crocker and Grier \cite{Crocker1994} and later modified by Sainis and coworkers \cite{Sainis2007}. With such an experiment, we deduce interparticle potentials for individual PMMA particle pairs in mixtures of CHB and decalin. To trap PMMA particles in the index matching solvent, we use an approach pioneered by Dullens and coworkers that consists in using core/shell particles with a refractive index matching shell and a higher refractive index core material. The cores can then be trapped by optical tweezers \cite{Liu2019}. In our case, PMMA colloids containing a polystyrene (PS) core are employed \cite{Klein2014}. Due to the refractive index mismatch of the core, no labelling of the the particles is necessary and their positions can be tracked with bright-field microscopy. Since the chosen core to shell volume ratio is 1:46, the particles used can be assumed to very closely mimic the behavior of pure PMMA colloids. With this system, we investigate the temporal changes of forces between pairs of colloids over the course of tens of days. While all particles initially are significantly charged, we find that the charging decreases by a factor of three within five days. In order to minimize the effect of charging, organic salts such as tetrabutylammonium chloride (TBAC) or bromide (TBAB) are commonly added to the dispersions \cite{Wu2009,Roller2020,Roller2021,Besseling2012}. This strategy is also employed for other systems similar to PMMA in CHB/decalin \cite{ElMasri2012,Zargar2013}. We therefore also investigated, how the addition of TBAB to a dispersion of PMMA particles in CHB/decalin affects the interparticle forces and found that already small amounts of TBAB lead to a significant hardening of the particle potentials. Temporal changes of the interparticle forces then become negligible. Yet, even at the highest TBAB concentrations tested, the potentials retained a non-negligible softness.

\section{Results and discussion}\label{sec2}

\subsection{Particle Synthesis}

\begin{figure*}
\includegraphics[width=\textwidth]{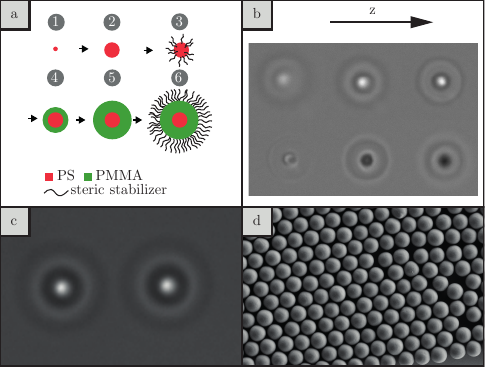}
\caption{(a) Particle synthesis (1) PS seeds synthesis in aqueous solution. (2) First and second PS shell growth. (3) Adhesion of a steric stabilizer and transfer to n-hexane/dodecane. (4, 5) PMMA shell growth. (6) Smoothing and covalent bonding of the steric stabilizer. (b) Bright field images of a particle when moved along the z-axis with an optical tweezer. While the PS core remains visible, the PMMA shell contrast vanishes when in focus. (c) Experimental bright field image. (d) SEM image of the core/shell particles. Scale bar: 5$\mu$m. }
\label{fig1}
\end{figure*}

The protocol for the synthesis of the PS/PMMA core/shell particles shown in Fig.~\ref{fig1} was adapted from Klein et al.\cite{Klein2014}. Following this procedure, first PS seed particles with a diameter of 190~nm were synthesized. Their diameter was increased to 600~nm by growing an additional PS layer in an emulsion polymerisation step. To this end, 35~ml doubly distilled water (Carl Roth) and 0.6~ml of the cores suspended in water (solid content 3.4\%) were heated to 73$^\circ$C under nitrogen in a 100~ml Schlenk flask. Meanwhile, 73~mg of potassium peroxodisulfate (Sigma Life Science) were mixed with 10~ml doubly distilled water and a monomer mixture consisting of 10~g distilled styrene (Merck) and 0.4~g 1,3-diisopropenylbenzene (DIPB, Tokyo Chemical Industries) was prepared. 1~ml of the K$_2$S$_2$O$_8$ mixture was given to the particles. 10~min later the nitrogen flow was stopped and 1.6~ml of the monomer mixture were added with a rate of 0.8~ml/h using a syringe pump. The particles were stirred for another two hours, then cooled down to room temperature, and filtered with glass wool. The same seeded emulsion polymerisation step was repeated with 4~ml of the synthesized particle solution and 32 ml of doubly destilled water  to create bigger PS cores with a diameter of 1.5~µm. These were then transferred to n-hexane/dodecane as described by Klein et al. \cite{Klein2014}. First, the particles were centrifuged and the supernatant was replaced with acetone. We found it necessary to change the duration of the particles being in acetone to at least 12~h. Then the particles were transferred to 19.5~g petroleum ether (PE) (boiling point $\geq$90\% 40-60$^\circ$C, Sigma-Aldrich). 30 drops of the steric stabilizer poly(12-hydroxystearic acid) (PHSA) grafted onto a backbone of PMMA (PHSA$-$\textit{g}$-$PMMA)\cite{Elsesser2010} were added. After 2~min of sonication, the particles were washed three times with PE and sonication in between before transfer to a mixture of n-hexane/dodecane (2:1 (wt\%:wt\%)) (n-hexane: for HPLC, VWR Chemicals). The particles were again centrifuged and filled up with 1.23~g of the n-hexane/dodecane solution.

With the PS cores as seed, a seeded  dispersion  polymerization step was used to grow a first PMMA shell resulting in particles with a diameter of 2.8~µm. For this purpose, a monomer mixture consisting of 21.3~ml methyl methacrylat (Sigma-Aldrich), 0.39~ml methacrylic acid (Sigma-Aldrich), 2.5~ml stabilizer, and 20.8~g n-hexane/dodecane mixture was prepared. In a 25~ml Schlenk tube, 39~mg  azo-bis-isobutyronitrile (Sigma-Aldrich) and the PS seed particles in n-hexane/dodecane were stirred with 250~rpm with a 1~cm magnetic stirrer bar. 8 $\mu$l octyl mercaptan (Sigma-Aldrich) were added and the particles were heated to 95$^\circ$C. 4.8~g of the monomer mixture were added at a rate of 3~ml/h using a syringe pump. After waiting for 2~h and cooling down to room temperature, the particles were filtered, and washed with PE. After the last centrifugation, the supernatant was removed and the particles were filled up with 2~ml PE. 0.3~ml of the particles were surface smoothed with a mixture of 9.7~g cis/trans-decalin ($>$98\%, Carl Roth) and 1.5~g acetone and 2 drops of stabilizer for 40~min \cite{Schuetter2017}. It was important to thoroughly mix the smoothing solution before adding the particles. After quenching with 25~ml decalin, the particles were then transferred to pure decalin. 0.29~g (solid weight) of these particles were used again as seed particles for another seeded dispersion polarization leading to particles with a diameter of 5.5~µm. This resulted in 1.6~g of particles. The particles were again smoothed with a mixture of 135~g decalin and 21~g acetone for 1.5~h before transfer to pure decalin. Finally, the particles were sterically stabilized by locking (PHSA$-$\textit{g}$-$PMMA) covalently onto the particles' surface. During smoothing, some particles coalesced. These were removed by sedimentation. The final particles had a core diameter of 1.42$\pm$0.08~µm (determined with SEM, 5.4\% polydispersity) and a shell diameter of 5.07$\pm 0.12$~µm (determined with SEM, 2.5\% polydispersity).

\subsection{Sample Preparation}

The solvent for the particles was a (85wt\%/15wt\%) mixture of CHB ($>$98\%, Tokyo Chemical Industries) and decalin. We only used CHB from freshly opened bottles, as the ion concentration in CHB is changing over time due to dissociation \cite{Green1955,Golinkin1970,Evans2016,Choi2019}. To avoid contact between particles, the particle concentration was kept smaller than 10 per $\mu$l, equivalent to a volume fraction of $\phi \approx {8\cdot10^{-7}}$. In some experiments, tetrabutylammonium bromide (TBAB, Sigma-Aldrich) was added to quantify the screening effect of this salt. In these cases, first a 368~$\mu$M solution of TBAB dissolved in CHB was prepared under constant stirring in nitrogen atmosphere over a minimum of three days. When necessary, this solution was diluted with CHB. Then the CHB/TBAB solution was mixed with decalin containing the particles. 

The sample chamber consisted of a glass slide (75~x~25~x~3~mm), with a centered round pit (diameter 8~mm, depth 0.5~mm) on the bottom side. From the other side, an additional small hole (diameter 2.8~mm, depth 2.5~mm) was cut. Before usage, these glass slides and glass coverslips (18~x~18~x~0.17~mm, Marienfeld) were washed for an hour in an ultrasonic bath set to 45°C first with doubly distilled water and then for another hour in ethanol (for spectroscopy, Uvasol\textsuperscript{\textregistered} Supelco\textsuperscript{\textregistered}) before drying in a nitrogen flow. Care was taken to touch the glasses only with clean tweezers. The bottom whole was covered with a coverslip and the edges were sealed with an epoxy resin (UHU Plus Sofortfest). As the resin is known to influence the interparticle potential when not properly hardened \cite{Choi2019}, we waited at least 12~h before filling in the sample liquid. After the liquid was inserted with a glass pipette, another glass coverslip was put on the top hole and sealed with epoxy resin. To avoid contact of the soft resin with the sample liquid, we waited for another 12~h before moving the sample chamber. 

\subsection{Experimental Setup and Raw Data Processing}

The experimental setup consists of a diode laser (P=250~mW, $\lambda=785$~nm,  FPL785S250, Thorlabs). The laser beam is directed over a spatial light modulator (SLM, X10468, Hamamatsu) and coupled via relay optics into a microscope (DMI 6000B, Leica). Phase masks were sent to the SLM to form two optical traps in the sample. Interparticle forces were deduced with a well-established method \cite{Sainis2007}. In brief, two particles were trapped and released periodically at initial distances $\Delta R$ between the particle centers. The laserdiode was switched on and off by a program every 500~ms. The off-times were 100~ms. During the time the traps were switched off, images were taken with an exposure time of 250~µs at a rate of 1000~fps using a CMOS camera (mvBlueFOX3-2 2004 C, Matrix Vision). Pixel sizes in the image plane were 115.5~nm x 115.5~nm. The free diffusing particles were recorded for 100 frames and from the resulting images, two dimensional trajectories $r_1 (t)$ and $r_2 (t)$ were generated. Typically, the statistics of a single force curve is based on more than $3 \times 10^4$ measurements. 

\begin{figure*}[htb]
\includegraphics[width=\textwidth]{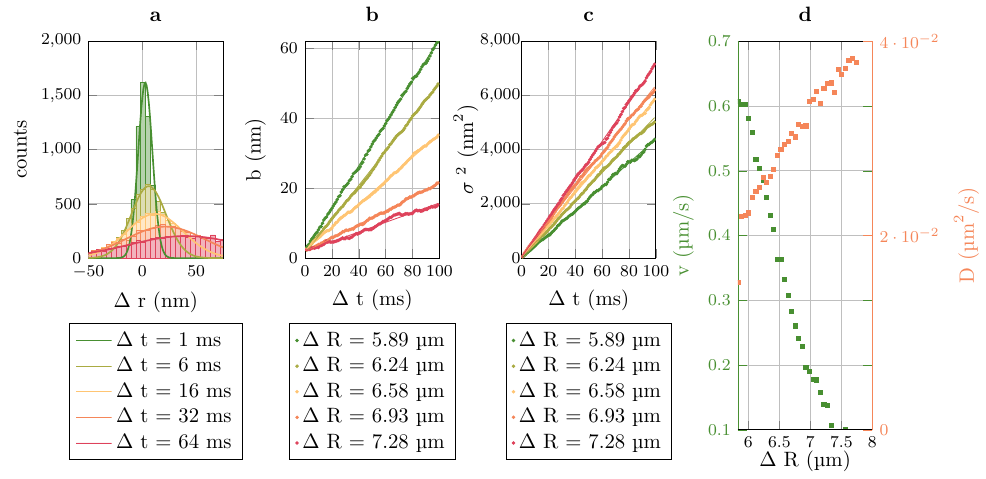}%
\caption{Raw data processing. The depicted data were taken from a sample without added TBAB . (a) Histogram of $\Delta r$, the change in particle separation after an interval $\Delta t$. For this histogram, the initial distance for all particle pairs was $\Delta R=$ 5.95~µm. The histogram well fitted by a Gaussian distribution. (b,c) Parameters $b$ and $\sigma^2$ of the Gaussian fits from (a). (d) The parameters $v$ and $D$, that are derived from the parameters $b$ and $c$. From $v$ and $D$, the force $F$ is calculated.}
\label{fig:DataProcessing}
\end{figure*}

To outline the data processing, exemplary data from samples without added TBAB are depicted in Fig.~\ref{fig:DataProcessing}. Fig.~\ref{fig:DataProcessing}a shows the measured distance distribution of $\Delta r =(r_1 (t)-r_2 (t))-(r_1 (t_0 )-r_2 (t_0))$ for several time intervals $\Delta t=t-t_0$. These curves are fitted with a Gaussian distribution function
$p(\Delta r)=A\cdot\exp((\Delta r-b)^2/(2\sigma^2))$.
The parameters $b$ and $\sigma^2$ are linearly dependent on  $\Delta t$ (Figure~\ref{fig:DataProcessing}b and Figure~\ref{fig:DataProcessing}c). A linear fit yields $v=(\mathrm{d} b)/(\mathrm{d} \Delta t)$ and $D=(\mathrm{d}  \sigma^2)/(2\mathrm{d} \Delta t)$ (Figure~\ref{fig:DataProcessing}d). With these parameters the forces are calculated using \cite{Sainis2007} 
$F=(k_B T\cdot v)/D$

The particles were slightly defocused during the measurements. This facilitates the fitting of the observed two dimensional intensity distribution using Gaussian functions (see Figure~\ref{fig1} b and c). In this manner, particle positions could be determined with subpixel precision. As the distance between measurement spot and glass surface might have an influence on the measured the potentials \cite{Hsu2005}, all data were acquired at depths 100~µm above the lower glass coverslip surface.

\subsection{Results}

\subsubsection{Temporal evolution of interparticle forces}

\begin{figure*}[htb]
\includegraphics[width=\textwidth]{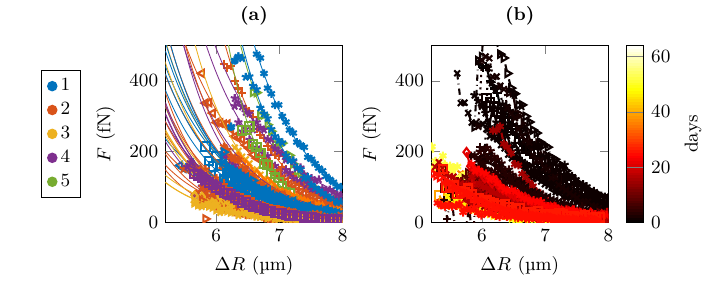}%
\caption{Interparticle forces determined from PMMA particles in CHB/decalin without added TBAB. (a) All data were recorded from different sample chambers plotted together. Color encodes for the different sample chambers used. (b) The same data as in (a) but now color coded for time after sample preparation.}
\label{fig:forces_chambers_time}
\end{figure*}

In a first series of experiments, we investigated samples of PMMA particles in CHB/decalin without added TBAB. Fig.~\ref{fig:forces_chambers_time}a shows all data recorded from different sample chambers plotted together. Since the force data are expected to be highly susceptible to impurities introduced during sample preparation, especially ions and glue residues in the sample fluid, we wondered whether the strong variations in the determined force curves reflected the purity of the samples. This reasoning was motivated by a previous report of similar variations in parameters like Debye lengths and effective charges for different samples despite their careful preparation \cite{vanderLinden2015}. Sorting of the same data as a function of time after sample preparation, however, shows that the observed variations are not caused by irregularities in the sample preparation, but are rather much due to a change of the samples as a function of time after preparation (cf. Fig.~\ref{fig:forces_chambers_time}b).

\begin{figure*}[htb]
\includegraphics[width=\textwidth]{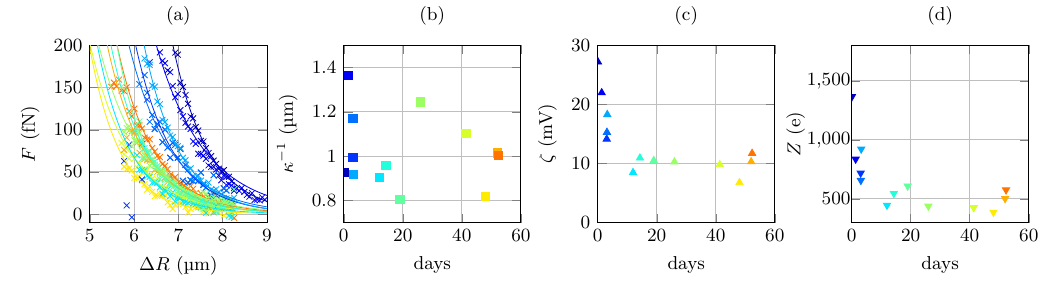}
\caption{Temporal evolution of the interparticle forces as a function of time after sample preparation. (a) Force curves from one sample cell without added TBAB. From the data, Debye lengths $\kappa^{-1}$ (b), surface potentials $\zeta$ (c), and surface charges Z (d) as a function of days after the preparation of the sample cell are derived.}
\label{fig:temporal_evolution}
\end{figure*}

This observation prompted us to analyze the data for a long series of experiments on one sample chamber in greater detail (Fig.~\ref{fig:temporal_evolution}). All the force measured in our experiments are well described by a screened Coulomb force \cite{Sainis2007}

\begin{equation}
F(r)= \frac{\left( 
e\zeta\right) ^2}{k_B T} \frac{a^2}{\lambda_b}
\quad \frac{\exp \left(-\kappa\left( r-2a\right) \right)}{r}\left(\frac{1}{r}+\kappa \right)
\label{eq:ScreenedCoulombForm}
\end{equation}

where $e$ is the elementary charge, $\zeta$ is the surface potential, $a$ is the particle radius, $\lambda_B=e^2/\left(4\pi\epsilon \epsilon_0k_B T)\right)$ is the Bjerrum length, and $\kappa^{-1}$ is the Debye screening length. The solid lines shown in the force data are fits with Eq.~(\ref{eq:ScreenedCoulombForm}), where $\kappa$ and $\zeta$ are fitting parameters. To account for particle polydispersity, the particle radius $a$, by contrast, was determined from bright-field microscopy. This was done by focusing the particles such that their outer rims became slightly visible. Knowing the Debye length $\kappa$ and the surface potential $\zeta$, an apparent surface charge can be calculated \cite{Hsu2005} 
\begin{equation}
Z= \frac {a \left( 1+\kappa a\right)} {\lambda_B} \frac{e \zeta}{k_B T}
\label{eq:Z}
\end{equation}

Our data show that the interaction between the colloids stabilizes roughly five days after sample preparation. The values for the determined Debye lengths are scattered between 0.8 and 1.4 $\mu m$, but show no trend in their temporal evolution. This means that the ion concentration in the dispersion fluid hardly changes with time. $\zeta$ potentials and, as a result also the apparent surface charges Z of the colloids, by contrast, clearly show a rapid decay immediately after sample preparation, before settling at constant values of approximately one third of their starting values after five days. The plateau values of the surface charges Z are on the order of 500 elementary charges.

It has been postulated that the charging of colloidal particles in CHB could be due to its dissociation and slightly preferential adsorption of protons \cite{vanderLinden2015,Choi2019}. Our method does not allow us to draw conclusions about the sign of the charges on the colloids, as $\zeta$ appears quadratically in Eq.~\ref{eq:ScreenedCoulombForm}. However, we can infer from the stable observed Debye lengths that in the samples investigated, no significant decomposition of CHB took place after their preparation. This does not exclude the possibility that a certain number of ions present in the CHB before filling the sample chambers is adsorbed by the particles. Also in this case, the degree of CHB dissociation must, however, be very small, since the dispersion contains only few particles that are able to adsorb the ions. A second and perhaps most likely possibility is that the colloids get charged during the filling procedure of the sample chambers. This charge would then be redistributed in the dispersion over the course of several days before reaching an equilibrium. The values for the various parameters that we determine after this equilibration process coincide well with earlier measurements of PMMA particles in CHB/\textit{cis}-decalin. Using microelectrophoresis, van der Linden and coworkers were able to determine apparent surface charges between 456 and 1015 elementary charges for slightly smaller particles with a diameter of 1.98 $\mu m$ \cite{vanderLinden2015}. Also there, strong variations in parameters like the apparent surface charge were reported in different samples despite identical preparation protocols. It is not known, whether also in their case differences in the sample preparation times account for this behavior. One should note, however, that the authors assumed a Debye length of $\kappa^{-1} = 6~\mu m$, significantly larger than our and other previously reported values for the same system \cite{Leunissen2005,Royall2006}.

\subsubsection{Influence of TBAB addition on interparticle forces}

As has been pointed out above, a common procedure to reach hard potentials in colloidal dispersions is to screen particle charges by the addition of salt. In organic dispersion media such as decalin used here, often TBAC or TBAB are employed\cite{Yethiraj2003}. We therefore also investigated the effect of TBAB on the forces between PMMA dispersed in CHB/decalin. 

\begin{figure*}[htb]
\includegraphics[width=\textwidth]{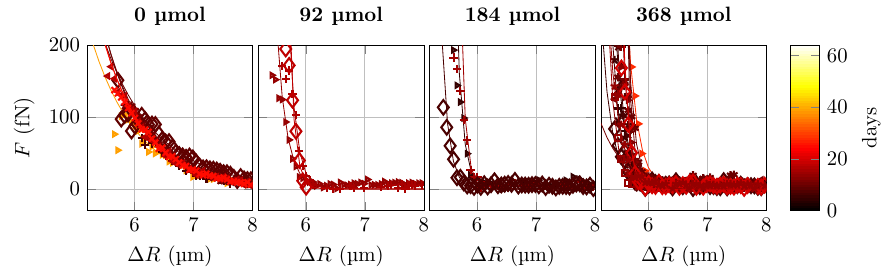}%
\caption{Interparticle forces as a function of particle separation $\Delta R$ for different concentration of TBAB. The solid lines are fits to the data according to a screened Coulomb force (Eq.~(\ref{eq:ScreenedCoulombForm})).}
\label{fig:ForceCurves}
\end{figure*}

Measurements for four different TBAB concentrations $c_{TBAB}$ (0~µmol, 92~µmol, 184~µmol and 368~µmol) are shown in Figure~\ref{fig:ForceCurves}. As discussed above, samples without added TBAB needed roughly five days to show a stable behaviour, therefore only measurements that were taken more than one week after filling are shown in Figure~\ref{fig:ForceCurves}. For each concentration, data for several colloid pairs were collected and each measurement shown is from a different particle pair. In the same sample cell, small differences between particle pairs are commonly observed. These can be explained by the polydispersity of the particles and by the previously observed phenomenon of fluctuating charges on colloids \cite{Strubbe2006,Strubbe2008,Schreuer2018}. In addition, variations in particle charges could also result from different degrees of the covalent binding of the steric stabilizer, that is known to strongly influence the surface charge \cite{vanderLinden2015}. 

The measured force curves show no sign of an attractive potential. This is expected due to the fact that firstly, the steric stabilisation of the PMMA particles reduces van~der~Waals forces and secondly, because our spatial resolution of 50~nm in $\Delta R$ is too small to observe the latter. The expected screening effect when adding TBAB is clearly visible from the shape of the force curves. Again, we determine values for the Debye lengths $\kappa^{-1}$, for $\zeta$potentials, and for apparent surface charges Z as described above. These values are collected in table~\ref{tab:FitparametersVariableR}.\\

\begin{table*}
\begin{tabular}{llllll}
\hline
$c_{TBAB}$  & $Nr_m$ & $ \zeta$ & $Z$ &$\kappa^{-1}$&$\phi_{contact}$\\
 (µ mol) &  & (mV) & ($e$)  &(µm)& ($k_B T$)\\
\hline
0 &  28&$10.6\pm2.3 $ & 527$\pm105 $  &$1.01\pm0.15$& 29.6\\
92 &  3 &$4.9\pm0.2 $ &  1541$\pm 495$ & 0.13$\pm 0.04$ & 6.0\\
184&   3& 2.4$\pm 0.1$ & 902$\pm167  $&$0.10\pm0.02 $& 1.5 \\
368&  13&  $1.8\pm 0.9$&  460$\pm138 $ &0.16$\pm 0.10$ & 1.0\\
\hline
\caption{\label{tab:FitparametersVariableR} Average fitting parameters from the fits shown in Fig.~\ref{fig:ForceCurves}. Uncertainties are standard deviations. All fits and also Z are calculated with a value for the particle radius a determined by light microscopy. $Nr_m$ is the number of measurements for a given TBAB concentration. $\phi_{contact}$ is the mean potential at contact, which was calculated by integration of Eq.~\ref{eq:ScreenedCoulombForm} and inserting a as determined by light microscopy.}
\end{tabular}
\end{table*}

The determined $\kappa^{-1}$ values indicate a clear change in the Debye length already when adding small concentrations of TBAB. Remarkably, the Debye length is roughly the same for all TBAB concentrations different from zero. This leads us to the conclusion, that above a concentration of 92 µmol TBAB, the free charge concentration in the sample liquid does not change much anymore, as the Debye length directly correlates with the number of free charges in the dispersion fluid. For $c_{TBAB}=0$~µmol, the value of $\kappa^{-1}=1.01$~µm is in the same range as the value of 1.4~µm for dilute systems found by Royall and coworkers \cite{Royall2006}, but differs from $\kappa^{-1}=12$~µm published earlier \cite{Yethiraj2003}. The results also agree well with the findings of Ref.~\citenum{Elbers2016}, where $\kappa^{-1}\approx1.2$~µm was determined for $c_{TBAB}=0$~µmol and $\kappa^{-1}\approx 0.2$~µm for $c_{TBAB}>50$~µmol in a system of 2.8~µm sized PMMA particles in CHB/cis-decalin. Similarly Leunissen and colleagues found $\kappa^{-1}=195$~nm for $c_{TBAB}$=190~µmol \cite{Leunissen2005}. 

The $\zeta$ potentials that we observed, decrease with higher TBAB concentrations. As expected, this means that the potential is getting harder even for larger concentrations $c_{TBAB}$, as for an ideal step function there should be a surface potential of zero. Also the apparent surface potential $\phi_{contact}$, which can be calculated from the integral of Eq.~\ref{eq:ScreenedCoulombForm}, shows a decrease from 29.6~$k_BT$ with no added TBAB to 1.0~$k_BT$ with $c_{TBAB}=$~368~µmol. These values are in good agreement with the results from Yethiraj and coworkers \cite{Yethiraj2003}, where crystallisation at a concentration of $c_{TBAB}=260$~µmol was observed to occur at a volume fraction $\phi=$0.42-0.45, indicating hard sphere like behaviour.

As pointed out, it is not possible for us to determine the sign of $\zeta$. Previous electrophoretic mobility measurements showed a change from positive to negative $\zeta$ potentials of the particles, when the TBAB concentration was added \cite{Royall2003,Leunissen2007,Kim2014}. We therefore assume that also in our case, the particle's $\zeta$ potentials changed from being positive without added TBAB to being negative for higher concentrations of TBAB. This means that we do not expect to observe a monotonous decrease of $\zeta$ as observed by Kim et al. \cite{Kim2014}. The absolute average charge $Z$ has its maximum value of 1541 elementary charges at $c_{TBAB}=92$~µmol. Since the apparent surface charge $Z$ is itself linearly dependent on $\zeta$, the sign of which cannot be determined with our method, also the sign of $\zeta$ remains unknown. Our values for Z compare well with experimental data reported earlier such as $Z = +221~e$ in CHB/decalin (27~wt\%) for 2.8~µm PMMA spheres \cite{Elbers2016}.

From the data depicted in Fig.~\ref{fig:ForceCurves} it is obvious that the effects from the addition of TBAB are so strong that in this case and in contrast to the samples made from pure CHB/decalin, no temporal evolution of the particle interaction is detected. This can be rationalized by the large number of ions present in the dispersion fluid after TBAB addition that overwhelms any small changes due to charge exchange between particles and fluid. 

\section{Conclusion}\label{sec13}

In this work, we report measurements of forces between individual pairs of PMMA particles dispersed in decalin. To measure these forces directly, we employed a variant of the blinking tweezer method \cite{Sainis2007}. The forces that we determine agree well with previously published data. As has been reported before by van der Linden and coworkers \cite{vanderLinden2015}, we also find, however, that the observed interparticle forces varied strongly. Using measurement series extending over tens of days, we could show that the variations are not due to impurities introduced during sample preparation, but to the temporal evolution of the particle charging. Since we found no concomitant temporal change of the Debye lengths, we assume that the charging of the particles is not caused by dissociation of CHB, but by the filling procedure. Quantitative measurements of the effect of the addition of TBAB to the dispersion as a screening agent shows that this increased the hardness of the particle potential. Already small amounts of TBAB lead to a large effect. In this case no temporal change of the interparticle forces was detected. However, even after TBAB addition, the particle potentials are not perfectly hard, as the contact potential even at the highest TBAB concentration was found to be approximately 1 $k_B T$. 

\section*{Acknowledgments}
The authors thank Franziska Rabold and John Geiger for help with the particle synthesis. This work was funded by the Deutsche Forschungsgemeinschaft (DFG) as SFB 1432, project C07.

\providecommand{\latin}[1]{#1}
\makeatletter
\providecommand{\doi}
  {\begingroup\let\do\@makeother\dospecials
  \catcode`\{=1 \catcode`\}=2 \doi@aux}
\providecommand{\doi@aux}[1]{\endgroup\texttt{#1}}
\makeatother
\providecommand*\mcitethebibliography{\thebibliography}
\csname @ifundefined\endcsname{endmcitethebibliography}
  {\let\endmcitethebibliography\endthebibliography}{}

\end{document}